\documentclass{article}
\usepackage{graphicx}
\usepackage{amsmath}

\input{tcilatex}

\begin{document}

\title{\textbf{High Energy Commutators in Particle, String and Membrane Theories}}
\author{W. Chagas-Filho \\
\textsl{Departamento de Fisica, Universidade Federal de Sergipe, Brazil }}
\maketitle

\begin{abstract}
We study relativistic particle, string and membrane theories as defining
field theories containing gravity in (0+1), (1+1) and (2+1) space-time
dimensions, respectively. We show how an off shell invariance of the
massless particle action allows the construction of an extension of the
conformal algebra and induces a transition to a non-commutative space-time
geometry. This non-commutative geometry is found to be preserved in the
space-time supersymmetric massless particle theory. It is then shown how the
basic bosonic commutators we found for the massless particle may also be
encountered in the tensionless limit of string and membrane theories.
Finally we speculate on how the non-locality introduced by these commutators
could be used to construct a covariant Newtonian gravitational field theory.
\end{abstract}

\section{Introduction}

\noindent Since some time ago there has been some interest in explaining a
very small measured value for the cosmological constant. It is believed that
this tinny value is a reflection of the presence of large amounts of \ dark
matter in the universe [1,2]. Because the measured value is positive, dark
matter should be contributing with positive gravitational potential energy
as a consequence of its interaction with usual matter. This implies that
dark matter should repel usual matter. We may add to this situation the
observation [3], using high resolution measurements of the cosmic microwave
background radiation, that the universe is effectively flat. On theoretical
grounds there is an interesting result obtained by Siegel [4], based in
scalar particle theory, that there are stringy corrections to gravitation
and that these corrections make the gravitational force less attractive at
very short distances. This work is an attempt to sketch a mathematical
picture in which all the above mentioned subjects would occupy
mathematically natural and consistent positions. The central mathematical
concept in this attempt will be non-commutativity of space-time coordinates

Non-commutativity is the central mathematical concept expressing uncertainty
in quantum mechanics, where it applies to any pair of conjugate variables.
But there are other situations where non-commutativity may manifest itself.
For instance, in the presence of a magnetic field even momenta fail to
mutually commute. The modern trend is to believe that one might postulate
non-commutativity for a number of reasons. One of these is the belief that
in quantum theories including gravity, space-time must change its nature at
the Planck scale. This is because quantum gravity has an uncertainty
principle which prevents one from measuring positions to better accuracies
than the Planck length. The momentum and energy required to make such a
measurement will itself modify the space-time geometry at this scale [5].
This effect could then be modeled by a non-vanishing commutation relation
between the space-time coordinates.

For most theories, postulating an uncertainty relation between space-time
coordinates will lead to a non-local theory and this may conflict with
Lorentz invariance. However, there is a remarkable exception to this
situation: string theory. String theory is not local in any sense we now
understand and indeed it has more than one parameter characterizing this
non-locality. One of these is the string length $l_{s}$, the average size of
a string. As we will see in this work, another useful parameter is the
string tension $T$ , the string analogue of the particle's mass. The string
tension defines an energy scale in string theory and the vanishing tension
limit is expected to describe the string dynamics at the Planck scale [6].

Historically, the first use of non-commutative geometry in string theory was
in Witten's formulation of open string field theory [7]. Witten's
formulation of non-commutative string field theory is based in the
introduction of a star product which is defined as the overlap of half of
two strings. The central idea in this formulation is to decompose the string
coordinates $x^{\mu }[\sigma ]$ as the direct sum $x^{\mu }[\sigma ]=l^{\mu
}[\sigma ]\oplus r^{\mu }[\sigma ]$ , where $l^{\mu }[\sigma ]$ and $r^{\mu
}[\sigma ]$ are respectively left and right coordinates of the string,
defined relative to the midpoint at $\sigma =\pi /2.$ If we now consider the
fields of two strings, say $A(l_{1}^{\mu }[\sigma ],r_{1}^{\mu }[\sigma ])$
and $B(l_{2}^{\mu }[\sigma ],r_{2}^{\mu }[\sigma ])$, then Witten's star
product gives 
\begin{equation}
A(l_{1}^{\mu }[\sigma ],r_{1}^{\mu }[\sigma ])\star B(l_{2}^{\mu }[\sigma
],r_{2}^{\mu }[\sigma ])=C(l_{1}^{\mu }[\sigma ],r_{2}^{\mu }[\sigma ]) 
\tag{1.1}
\end{equation}
where the field $C(l_{1}^{\mu }[\sigma ],r_{2}^{\mu }[\sigma ])$ is given by
the functional integral 
\begin{equation}
C(l_{1}^{\mu }[\sigma ],r_{2}^{\mu }[\sigma ])=\int [dz]A(l_{1}^{\mu
}[\sigma ],z^{\mu }[\sigma ])B(z^{\mu }[\sigma ],r_{2}^{\mu }[\sigma ]) 
\tag{1.2}
\end{equation}
and $z^{\mu }[\sigma ]=r_{1}^{\mu }[\sigma ]=l_{2}^{\mu }[\sigma ]$
corresponds to the overlap of half of the first string with half of the
second string.

Witten's formulation of open bosonic string field theory has experienced a
rebirth through new physical insights and technical advances. In particular,
it was shown by Bars [8] that Witten's star product, originally defined as a
path integral that saws two strings into a third one, is equivalent to the
Moyal star product [9], which is the central mathematical concept in the
familiar formulation of non-commutative geometry. But this equivalence is
rather subtle. As shown in [8], Witten's star product is equivalent to the
usual Moyal star product in the relativistic phase space of only the even
modes of oscillation of the string. This is because the mapping of the
Witten $\star $ to the Moyal $\star $ necessarily requires that the Fourier
space for the odd positions be named as the even momenta since $%
(x_{e},p_{e}) $ are canonical under the Moyal star-product. These
developments, and many others that have recently appeared in the literature,
led to the belief that non-commutative geometry naturally arises in string
field theory.

However, because it was defined for applications in traditional quantum
mechanics, the Moyal star product leaves unchanged the Heisenberg
commutation relations between the canonical operators. In the usual
formulations of field theory on non-commutative spaces, the position
coordinates satisfy commutation relations of the form [10,11] 
\begin{equation}
\lbrack q^{\mu },q^{\nu }]=i\theta ^{\mu \nu }  \tag{1.3}
\end{equation}
where $\theta ^{\mu \nu }$ is a constant anti-symmetric tensor. The effect
of such a modification is reflected in the momentum space vertices of the
theory by factors of the form [12] 
\begin{equation}
\exp \{i\theta ^{\mu \nu }p_{\mu }q_{\nu }\}  \tag{1.4}
\end{equation}
The first quantized string in the light cone gauge, as a perturbative
non-commu tative theory based in a commutation relation such as (1.3), was
studied in [12]. It was found by the authors in [12] that, apart from
trivial factors, the structure of the non-commutative theory is identical to
the structure of the commutative one. It thus seems that nothing
substantially new is introduced in string theory by the use of a Moyal
star-product structure, or by postulating another star-product structure
based on commutators such as (1.3).

In this work we give a small contribution to this subject by showing how we
may use the special-relativistic ortogonality between velocity and
acceleration to induce the appearance of a new invariance of the massless
relativistic particle action. This invariance may then be used to generate
transitions to new space-time coordinates which, in the simplest case,
satisfy commutator relations of the form 
\begin{equation}
\lbrack x_{\mu },x_{\nu }]\sim iM_{\mu \nu }^{\ast }  \tag{1.5}
\end{equation}
where $M_{\mu \nu }^{\ast }$ is an off shell anti-symmetric operator that
generates space-time rotations. This commutation relation is part of a
modified Heisenberg commutator structure which reduces to the usual
canonical commutators when the mass shell condition is imposed. Most part of
this work is the explicit verification that the basic commutator structure
we found for the massless particle may also be encountered in string and
membrane theory. As pointed out in [11], any finite-dimensional deformed
Poisson structure can be canonically quantized. It may be then interesting
to investigate what quantum effects we may find in particle, string and
membrane theories if we construct the corresponding non-commutative theories
based on commutators such as (1.5) instead of (1.3).

An important point to our considerations here is that space-time
non-commu-tativity introduces an associated non-locality [10]. In
non-commutative gauge theories based on the commutator (1.3) this
non-locality allows the interpretation of the gauge particles as dipoles
carrying opposite charges of the corresponding gauge theory [10, 12]. In the
context of string theory, it was shown [13] that an open string with both
ends on a (D2-D0)-brane system in which a magnetic field is defined looks
like an electric dipole if a background electric field is added. The
dipole-background interaction gives the flux modifications needed for the
Born-Infeld action on the non-commutative torus and results in a Hamiltonian
that depends on two integers, $n$ and $m$, whose ratio $\frac{m}{n}$ gives
the density of D0-branes distributed on the D2-brane [13].

In this work we study relativistic particle, string and membrane theories as
gauge systems in which the gauge invariance is general covariance. We
interpret these relativistic objects as defining field theories containing
gravity in (0+1), (1+1) and (2+1) space-time dimensions, respectively. The
paper is organized as follows: For the task of clarity, in section two we
briefly review the concept of a space-time conformal vector field and the
associated conformal algebra. In section three we discuss particle theory
and show how we may induce an off shell invariance of the massless particle
action and how this invariance allows the construction of an off shell
extension of the conformal algebra and how it permits a transition to
non-commutative space-time coordinates. In section four we show that the
bosonic commutator structure we found for the massless particle is preserved
in the space-time supersymmetric extension of the theory. Section five deals
with bosonic strings. We present an alternative string action that in
principle does not require a critical dimension and that allows a smooth
transition to the tensionless limit of string theory. We show how this
alternative string action may be gauge-fixed to yield a non-commutative
theory at the unitary tension value and to give the extension of the
particle commutators in the vanishing tension limit. Section six extends
these results to superstring theory and section seven briefly describes the
same situation in relativistic membrane theory. We present our conclusions
in section eight, where we also speculate on the possibility of using the
non-locality introduced by our commutation relations to provide the
conceptual basis for a covariant Newtonian gravitational field theory.

\section{Conformal Vector Fields}

Consider the Euclidean flat space-time vector field 
\begin{equation}
\hat{R}(\epsilon )=\epsilon ^{\mu }\partial _{\mu }  \tag{2.1}
\end{equation}
such that 
\begin{equation}
\partial _{\mu }\epsilon _{\nu }+\partial _{\nu }\epsilon _{\mu }=\delta
_{\mu \nu }\partial .\epsilon  \tag{2.2}
\end{equation}
The vector field $\hat{R}(\epsilon )$ gives rise to the coordinate
transformation 
\begin{equation}
\delta x^{\mu }=\hat{R}(\epsilon )x^{\mu }=\epsilon ^{\mu }  \tag{2.3}
\end{equation}
The vector field (2.1) is known as the Killing vector field and $\epsilon
^{\mu }$ is known as the Killing vector. One can show that the most general
solution for equation (2.2) in a four-dimensional space-time is 
\begin{equation}
\epsilon ^{\mu }=\delta x^{\mu }=a^{\mu }+\omega ^{\mu \nu }x_{\nu }+\alpha
x^{\mu }+(2x^{\mu }x^{\nu }-\delta ^{\mu \nu }x^{2})b_{\nu }  \tag{2.4}
\end{equation}
The vector field $\hat{R}(\epsilon )$ for the solution (2.4) can then be
written as 
\begin{equation}
\hat{R}(\epsilon )=a^{\mu }P_{\mu }-\frac{1}{2}\omega ^{\mu \nu }M_{\mu \nu
}+\alpha D+b^{\mu }K_{\mu }  \tag{2.5}
\end{equation}
where 
\begin{equation}
P_{\mu }=\partial _{\mu }  \tag{2.6}
\end{equation}
\begin{equation}
M_{\mu \nu }=x_{\mu }\partial _{\nu }-x_{\nu }\partial _{\mu }  \tag{2.7}
\end{equation}
\begin{equation}
D=x^{\mu }\partial _{\mu }  \tag{2.8}
\end{equation}
\begin{equation}
K_{\mu }=(2x_{\mu }x^{\nu }-\delta _{\mu }^{\nu }x^{2})\partial _{\nu } 
\tag{2.9}
\end{equation}
$P_{\mu }$ generates translations, $M_{\mu \nu }$ generates rotations, $D$
generates dilatations and $K_{\mu }$ generates conformal transformations in
space-time. The generators of the vector field $\hat{R}(\epsilon )$ obey the
commutator algebra 
\begin{equation}
\lbrack P_{\mu },P_{\nu }]=0  \tag{2.10a}
\end{equation}
\begin{equation}
\lbrack P_{\mu },M_{\nu \lambda }]=\delta _{\mu \nu }P_{\lambda }-\delta
_{\mu \lambda }P_{\nu }  \tag{2.10b}
\end{equation}
\begin{equation}
\lbrack M_{\mu \nu },M_{\lambda \rho }]=\delta _{\nu \lambda }M_{\mu \rho
}+\delta _{\mu \rho }M_{\nu \lambda }-\delta _{\nu \rho }M_{\mu \lambda
}-\delta _{\mu \lambda }M_{\nu \rho }  \tag{2.10c}
\end{equation}
\begin{equation}
\lbrack D,D]=0  \tag{2.10d}
\end{equation}
\begin{equation}
\lbrack D,P_{\mu }]=-P_{\mu }  \tag{2.10e}
\end{equation}
\begin{equation}
\lbrack D,M_{\mu \nu }]=0  \tag{2.10f}
\end{equation}
\begin{equation}
\lbrack D,K_{\mu }]=K_{\mu }  \tag{2.10g}
\end{equation}
\begin{equation}
\lbrack P_{\mu },K_{\nu }]=2(\delta _{\mu \nu }D-M_{\mu \nu })  \tag{2.10h}
\end{equation}
\begin{equation}
\lbrack M_{\mu \nu },K_{\lambda }]=\delta _{\nu \lambda }K_{\mu }-\delta
_{\lambda \mu }K_{\nu }  \tag{2.10i}
\end{equation}
\begin{equation}
\lbrack K_{\mu },K_{\nu }]=0  \tag{2.10j}
\end{equation}
The commutator algebra (2.10) is the conformal space-time algebra in four
dimensions. Notice that the commutators (2.10a-2.10c) correspond to the
Poincar\'{e} algebra. The Poincar\'{e} algebra is a sub-algebra of the
conformal algebra. Let us now see how we can extend the conformal algebra,
and how this extended algebra is related to a non-commutative space-time
geometry.

\section{Relativistic Particles}

A relativistic particle describes is space-time a one-parameter trajectory $%
x^{\mu }(\tau )$. The dynamics of the particle must be independent of the
parameter choice. A possible form of the action is the one proportional to
the arc length traveled by the particle and given by 
\begin{equation}
S=-m\int ds=-m\int d\tau \sqrt{-\dot{x}^{2}}  \tag{3.1}
\end{equation}
In this work we take the parameter $\tau $ to be the particle's proper time, 
$m$ is the particle's mass and $ds^{2}=-\delta _{\mu \nu }dx^{\mu }dx^{\nu }$%
. A dot denotes derivatives with respect to $\tau $ and we use units in
which $\hbar =c=1$.

Action (3.1) is obviously inadequate to study the massless limit of the
theory and so we must find an alternative action. Such an action can be
easily computed by treating the relativistic particle as a constrained
system. In the transition to the Hamiltonian formalism action (3.1) gives
the canonical momentum 
\begin{equation}
p_{\mu }=\frac{m}{\sqrt{-\dot{x}^{2}}}\dot{x}_{\mu }  \tag{3.2}
\end{equation}
and this momentum gives rise to the primary constraint 
\begin{equation}
\phi =\frac{1}{2}(p^{2}+m^{2})=0  \tag{3.3}
\end{equation}
We follow Dirac's [14] convention that a constraint is set equal to zero
only after all calculations have been performed. The canonical Hamiltonian
corresponding to action (3.1), $H=p.\dot{x}-L$, identically vanishes. This
is a characteristic feature of reparametrization-invariant systems. Dirac's
Hamiltonian for the relativistic particle is then 
\begin{equation}
H_{D}=H+\lambda \phi =\frac{1}{2}\lambda (p^{2}+m^{2})  \tag{3.4}
\end{equation}
where $\lambda (\tau )$ is a Lagrange multiplier. The Lagrangian that
corresponds to (3.4) is 
\begin{eqnarray}
L &=&p.\dot{x}-H_{D}  \notag \\
&=&p.\dot{x}-\frac{1}{2}\lambda (p^{2}+m^{2})  \TCItag{3.5}
\end{eqnarray}
Solving the equation of motion for $\ p_{\mu }$ that \ follows from (3.5)
and inserting the result back in it, we obtain the particle action 
\begin{equation}
S=\int d\tau (\frac{1}{2}\lambda ^{-1}\dot{x}^{2}-\frac{1}{2}\lambda m^{2}) 
\tag{3.6}
\end{equation}
The great advantage of action (3.6) is that it has a smooth transition to
the $m=0$ limit.

Action (3.6) is invariant under the Poincar\'{e} transformations 
\begin{equation}
\delta x^{\mu }=a^{\mu }+\omega _{\nu }^{\mu }x^{\nu }  \tag{3.7a}
\end{equation}
\begin{equation}
\delta \lambda =0  \tag{3.7b}
\end{equation}
Invariance of action (3.6) under transformation (3.7a) implies that we can
construct a space-time vector field corresponding to the first two
generators in the right of equation (2.5). These generators realize the
Poincar\'{e} algebra (2.10a-2.10c).

Now we make a transition to the massless limit. This limit is described by
the action 
\begin{equation}
S=\frac{1}{2}\int d\tau \lambda ^{-1}\dot{x}^{2}  \tag{3.8}
\end{equation}
The massless limit is usually considered to describe the high energy limit
of \ relativistic particle theory. The canonical momentum conjugate to $%
x^{\mu }$ is 
\begin{equation}
p_{\mu }=\frac{1}{\lambda }\dot{x}_{\mu }  \tag{3.9}
\end{equation}
The canonical momentum conjugate to $\lambda $ identically vanishes and this
is a primary constraint, $p_{\lambda }=0$. Constructing the canonical
Hamiltonian, and requiring the stability of this constraint, we are led to
the mass shell condition 
\begin{equation}
\phi =\frac{1}{2}p^{2}=0  \tag{3.10}
\end{equation}
Let us now study which space-time symmetries are present in this limit.
Being the $m=0$ limit of (3.6), action (3.8) is also invariant under the
Poincar\'{e} transformation (3.7). The massless action (3.8) however has a
larger set of space-time invariances. It is also invariant under the scale
transformation 
\begin{equation}
\delta x^{\mu }=\alpha x^{\mu }  \tag{3.11a}
\end{equation}
\begin{equation}
\delta \lambda =2\alpha \lambda  \tag{3.11b}
\end{equation}
where $\alpha $ is a constant, and under the conformal transformation 
\begin{equation}
\delta x^{\mu }=(2x^{\mu }x^{\nu }-\delta ^{\mu \nu }x^{2})b_{\nu } 
\tag{3.12a}
\end{equation}
\begin{equation}
\delta \lambda =4\lambda x.b  \tag{3.12b}
\end{equation}
where $b_{\mu }$ is a constant vector. Invariance of action (3.8) under
transformations (3.7a), (3.11a) and (3.12a) then implies that the full
conformal field (2.5) can be defined in the massless sector of the theory.

It is convenient at this point to relax the mass shell condition (3.10)
because the massless particle will be off mass shell in the presence of
interactions [15]. We may now use the fact that we are dealing with a
special-relativistic system. Special relativity has the characteristic
kinematical feature that the relativistic velocity is always ortogonal to
the relativistic acceleration ( see, for instance, [16] ). While the
mathematics involved at this point is very simple, the physical implications
of this ortogonality condition are rather unexplored. In this work we make
an attempt to partially clarify these physical implications in the context
of massless relativistic particle theory. We start by adopting the point of
view that special relativity automatically implies the ortogonality
condition between the velocity and the acceleration. Reasoning in this way,
we find that we may use this ortogonality to induce the presence of a new
invariance of the massless particle action. Then, as a consequence of the
fact that action (3.8) is a special-relativistic action, it is invariant
under the transformation 
\begin{equation}
x^{\mu }\rightarrow \tilde{x}^{\mu }=\exp \{\beta (\dot{x}^{2})\}x^{\mu } 
\tag{3.13a}
\end{equation}
\begin{equation}
\lambda \rightarrow \exp \{2\beta (\dot{x}^{2})\}\lambda  \tag{3.13b}
\end{equation}
where $\beta $ is an arbitrary function of $\dot{x}^{2}$. We emphasize that
although the ortogonality condition must be used to get the invariance of
action (3.8) under transformation (3.13), this condition is not an external
ingredient in the theory. In fact, the ortogonality between the relativistic
velocity and the acceleration is an unavoidable condition here, it is an
imposition of special relativity. We are just using this aspect of special
relativity to get an invariant action.

Now, invariance of the massless action under transformations (3.13) means
that infinitesimally we can define the scale transformation $\delta x^{\mu
}=\alpha \beta (\dot{x}^{2})x^{\mu }$ , where $\alpha $ is the same constant
that appears in equations (2.4) and (2.5). These transformations then lead
to the existence of a new type of dilatations. These new dilatations
manifest themselves in the fact that the vector field $D$ of equation (2.8)
can be changed according to 
\begin{equation}
D=x^{\mu }\partial _{\mu }\rightarrow D^{\ast }=x^{\mu }\partial _{\mu
}+\beta (\dot{x}^{2})x^{\mu }\partial _{\mu }=D+\beta D  \tag{3.14}
\end{equation}
In fact, because all vector fields in equation (2.5) involve partial
derivatives with respect to $x^{\mu }$ and $\beta $ is a function of $\dot{x}%
^{\mu }$ only, we can also introduce the generators 
\begin{equation}
P_{\mu }^{\ast }=P_{\mu }+\beta P_{\mu }  \tag{3.15}
\end{equation}
\begin{equation}
M_{\mu \nu }^{\ast }=M_{\mu \nu }+\beta M_{\mu \nu }  \tag{3.16}
\end{equation}
\begin{equation}
K_{\mu }^{\ast }=K_{\mu }+\beta K_{\mu }  \tag{3.17}
\end{equation}
and define the new space-time vector field 
\begin{equation}
V_{0}^{\ast }=a^{\mu }P_{\mu }^{\ast }-\frac{1}{2}\omega ^{\mu \nu }M_{\mu
\nu }^{\ast }+\alpha D^{\ast }+b^{\mu }K_{\mu }^{\ast }  \tag{3.18}
\end{equation}
The generators of this vector field obey the algebra 
\begin{equation}
\lbrack P_{\mu }^{\ast },P_{\nu }^{\ast }]=0  \tag{3.19a}
\end{equation}
\begin{equation}
\lbrack P_{\mu }^{\ast },M_{\nu \lambda }^{\ast }]=(\delta _{\mu \nu
}P_{\lambda }^{\ast }-\delta _{\mu \lambda }P_{\nu }^{\ast })+\beta (\delta
_{\mu \nu }P_{\lambda }^{\ast }-\delta _{\mu \lambda }P_{\nu }^{\ast }) 
\tag{3.19b}
\end{equation}
\begin{equation*}
\lbrack M_{\mu \nu }^{\ast },M_{\lambda \rho }^{\ast }]=(\delta _{\nu
\lambda }M_{\mu \rho }^{\ast }+\delta _{\mu \rho }M_{\nu \lambda }^{\ast
}-\delta _{\nu \rho }M_{\mu \lambda }^{\ast }-\delta _{\mu \lambda }M_{\nu
\rho }^{\ast })
\end{equation*}
\begin{equation}
+\beta (\delta _{\nu \lambda }M_{\mu \rho }^{\ast }+\delta _{\mu \rho
}M_{\nu \lambda }^{\ast }-\delta _{\nu \rho }M_{\mu \lambda }^{\ast }-\delta
_{\mu \lambda }M_{\nu \rho }^{\ast })  \tag{3.19c}
\end{equation}
\begin{equation}
\lbrack D^{\ast },D^{\ast }]=0  \tag{3.19d}
\end{equation}
\begin{equation}
\lbrack D^{\ast },P_{\mu }^{\ast }]=-P_{\mu }^{\ast }-\beta P_{\mu }^{\ast }
\tag{3.19e}
\end{equation}
\begin{equation}
\lbrack D^{\ast },M_{\mu \nu }^{\ast }]=0  \tag{3.19f}
\end{equation}
\begin{equation}
\lbrack D^{\ast },K_{\mu }^{\ast }]=K_{\mu }^{\ast }+\beta K_{\mu }^{\ast } 
\tag{3.19g}
\end{equation}
\begin{equation}
\lbrack P_{\mu }^{\ast },K_{\nu }^{\ast }]=2(\delta _{\mu \nu }D^{\ast
}-M_{\mu \nu }^{\ast })+2\beta (\delta _{\mu \nu }D^{\ast }-M_{\mu \nu
}^{\ast })  \tag{3.19h}
\end{equation}
\begin{equation}
\lbrack M_{\mu \nu }^{\ast },K_{\lambda }^{\ast }]=(\delta _{\lambda \nu
}K_{\mu }^{\ast }-\delta _{\lambda \mu }K_{\nu }^{\ast })+\beta (\delta
_{\lambda \nu }K_{\mu }^{\ast }-\delta _{\lambda \mu }K_{\nu }^{\ast }) 
\tag{3.19i}
\end{equation}
\begin{equation}
\lbrack K_{\mu }^{\ast },K_{\nu }^{\ast }]=0  \tag{3.19j}
\end{equation}
Notice that the vanishing brackets of the conformal algebra (2.10) are
preserved as vanishing in the above algebra, but the non-vanishing brackets
of the conformal algebra now have linear and quadratic contributions from
the arbitrary function $\beta (\dot{x}^{2})$. Algebra (3.19) is an off shell
extension of the conformal algebra (2.10).

Now consider the commutator structure induced by transformation (3.13a). We
assume the usual commutation relations between the canonical variables, $%
[x_{\mu },x_{\nu }]=[p_{\mu },p_{\nu }]=0$ , $[x_{\mu },p_{\nu }]=i\delta
_{\mu \nu }$. Taking $\beta (\dot{x}^{2})=\beta (\lambda ^{2}p^{2})$ in
transformation (3.13a) and transforming the $p_{\mu }$ in the same manner as
the $x_{\mu },$ we find that the new transformed canonical variables $(%
\tilde{x}_{\mu },\tilde{p}_{\mu })$ obey the commutators 
\begin{equation}
\lbrack \tilde{p}_{\mu },\tilde{p}_{\nu }]=0  \tag{3.20}
\end{equation}
\begin{equation}
\lbrack \tilde{x}_{\mu },\tilde{p}_{\nu }]=i\delta _{\mu \nu }(1+\beta
)^{2}+(1+\beta )[x_{\mu },\beta ]p_{\nu }  \tag{3.21}
\end{equation}
\begin{equation}
\lbrack \tilde{x}_{\mu },\tilde{x}_{\nu }]=(1+\beta )\{x_{\mu }[\beta
,x_{\nu }]-x_{\nu }[\beta ,x_{\mu }]\}  \tag{3.22}
\end{equation}
written in terms of the old canonical variables. These commutators obey the
non trivial Jacobi identities $(\tilde{x}_{\mu },\tilde{x}_{\nu },\tilde{x}%
_{\lambda })=0$ \ and \ $(\tilde{x}_{\mu },\tilde{x}_{\nu },\tilde{p}%
_{\lambda })=0$. They also reduce to the usual canonical commutators when $%
\beta (\lambda ^{2}p^{2})=0$.

It may be questioned here, on the basis of the definition (3.9) of the
classical canonical momentum, if the quantum momentum should not transform
in the inverse way of the $x^{\mu }$ under (3.13), as is the case for the
classical momemtum. This is an interesting point in the mathematical
developments here. In fact, the identification of the correct quantum
canonical momentum is a typical problem in field theories defined over a
non-commutative space-time [24]. An example is non-commutative quantum
string field theory, where the momentum space Fourier transform of the odd
vibrational position modes behave as the quantum canonical momenta conjugate
to the even vibrational position modes [8]. In the context of this work,
transforming the quantum momenta as the classical ones implies substituting
the commutator (3.21) by the commutator 
\begin{equation*}
\lbrack \tilde{x}_{\mu },\tilde{p}_{\nu }]=(1+\beta )\{i\delta _{\mu \nu
}(1-\beta )-[x_{\mu },\beta ]p_{\nu }\}
\end{equation*}
This commutator also reduces to the usual Heisenberg commutator when $\beta
(\lambda ^{2}p^{2})=0$. However, the Jacobi identity $(\tilde{x}_{\mu },%
\tilde{x}_{\nu },\tilde{p}_{\lambda })=0,$ based on commutator (3.22) and on
the above commutator does not seem to close. If this is the case, the
classical and quantum canonical momenta of the massless particle on a
non-commutative space-time are related by a scale transformation of the type
(3.13). That is, $p_{q}=\exp \{2\beta \}p_{c}$.

The simplest example of non-commutative space-time geometry induced by
transformation (3.13) is the case when $\beta (\lambda ^{2}p^{2})=\lambda
^{2}p^{2}$. The new positions then satisfy 
\begin{equation}
\lbrack \tilde{x}_{\mu },\tilde{x}_{\nu }]=-2i\lambda ^{2}M_{\mu \nu }^{\ast
}  \tag{3.23}
\end{equation}
where $M_{\mu \nu }^{\ast }$ is the extended off shell operator of Lorentz
rotations given by equation (3.16). The commutator (3.23) satisfies 
\begin{equation}
\int \mathbf{Tr}[\tilde{x}_{\mu },\tilde{x}_{\nu }]=0  \tag{3.24}
\end{equation}
as is the case for a general non-commutative algebra [10]. From equation
(3.23) we may say that there exists an inertial frame in which the
uncertainty introduced by two simultaneous position measurements is an off
shell rotation in space-time. This appears to confirm the observation in
[25] that the non-commutative space-time is to be interpreted as having an
internal angular momentum throughout.

We may now consider the question of if we can find a mathematical structure
that would behave like a dipole in the context of massless particle theory.
Such a mathematical structure may be constructed by considering the most
general relativistic particle of mass $m$ that special relativity allows us
to construct. This corresponds to a particle with an internal structure
formed of positive and negative multiples of a fundamental mass $\mu $ ,
that is, 
\begin{equation*}
m=(n_{+}+n_{-})\mu
\end{equation*}
with $n_{+}=1,2,...$ and $n_{-}=-1,-2,...$ The values $n_{+}=n_{-}=0$ are
not allowed here because we interpret the fundamental mass $\mu $ as a
reflection of the existence of a fundamental length. The reader should
recall at this point that special relativity allows the existence of
negative masses trough the equation 
\begin{equation*}
E^{2}=p^{2}+m^{2}
\end{equation*}
(in our system of units). We may then interpret the non-locality in
space-time introduced by the commutator (3.23) as an indication that the
massless relativistic particle is actually a gravitational dipole with $%
n_{+}=1$ and $n_{-}=-1$.

\section{Superparticles}

In this section we consider the case of the massless superparticle. The
bosonic massless action (3.8) has the supersymmetric extension [17] 
\begin{equation}
S=\frac{1}{2}\int d\tau \lambda ^{-1}(\dot{x}^{\mu }-i\bar{\theta}\Gamma
^{\mu }\dot{\theta})^{2}  \tag{4.1}
\end{equation}
where $\theta _{\alpha }$ is a space-time spinor and $\Gamma _{\alpha \beta
}^{\mu }$ are Dirac matrices. Action (4.1) is invariant under the global
supersymmetry transformation 
\begin{equation}
\delta x^{\mu }=i\bar{\epsilon}\Gamma ^{\mu }\theta  \tag{4.2a}
\end{equation}
\begin{equation}
\delta \theta =\epsilon  \tag{4.2b}
\end{equation}
\begin{equation}
\delta \lambda =0  \tag{4.2c}
\end{equation}
where $\epsilon $ is an infinitesimal constant Grassmann parameter. The
canonical momenta that follow from action (4.1) are 
\begin{equation}
p_{\lambda }=0  \tag{4.3}
\end{equation}
\begin{equation}
p_{\mu }=\frac{1}{\lambda }(\dot{x}_{\mu }-i\bar{\theta}\Gamma _{\mu }\dot{%
\theta})  \tag{4.4}
\end{equation}
\begin{equation}
\pi _{\alpha }=i(p.\Gamma \theta )_{\alpha }  \tag{4.5}
\end{equation}
Equation (4.5) leads to the fermionic constraint $\pi _{\alpha }-i(p.\Gamma
\theta )_{\alpha }=0.$ As in the bosonic case, requiring the stability of
constraint (4.3), we are led to the constraint 
\begin{equation}
\phi =\frac{1}{2}p^{2}=0  \tag{4.6}
\end{equation}
which is the mass shell condition for the superparticle. It is convenient to
introduce the variable $Z_{0}^{\mu }=\dot{x}^{\mu }-i\bar{\theta}\Gamma
^{\mu }\dot{\theta}$. This variable is invariant under transformation (4.2)
all by itself \ and so all manipulations involving $Z_{0}^{\mu }$ will
automatically be supersymmetric invariant.

If we extend the calculations in [16] to the case of the massless
superparticle, we will find that the relation 
\begin{equation}
Z_{0}.\frac{dZ_{0}}{d\tau }=0  \tag{4.7}
\end{equation}
must hold. Equation (4.7) is the supersymmetric extension of the
special-relativistic condition of ortogonality between velocity and
acceleration. Again, this is an imposition of the supersymmetric
relativistic theory, and not an artificially introduced external ingredient.
Relaxing again the mass shell condition, we find that action (4.1) is
invariant under the transformation 
\begin{equation}
x^{\mu }\rightarrow \tilde{x}^{\mu }=\exp \{\beta (Z_{0}^{2})\}x^{\mu } 
\tag{4.8a}
\end{equation}
\begin{equation}
\theta _{\alpha }\rightarrow \tilde{\theta}_{\alpha }=\exp \{\frac{1}{2}%
\beta (Z_{0}^{2})\}\theta _{\alpha }  \tag{4.8b}
\end{equation}
\begin{equation}
\lambda \rightarrow \exp \{2\beta (Z_{0}^{2})\}\lambda  \tag{4.8c}
\end{equation}
where $\beta $ is now an arbitrary function of $Z_{0}^{2}.$ In the canonical
formalism $\beta (Z_{0}^{2})=\beta (\lambda ^{2}p^{2}).$ Since the bosonic
momentum $p^{\mu }$ commutes with the fermionic canonical variables $\theta
_{\alpha }$ and $\pi _{\alpha }$ , this leaves invariant the
anti-commutation relations between the fermi-onic variables but change the
bosonic ones in the same way as (3.20-3.22). The commutator structure we
found for the bosonic massless particle is thus preserved in the
supersymmetric massless theory.

\section{Relativistic strings}

Strings are higher-dimensional extensions of the particle concept. As a
consequence of its evolution, the string traces out a world sheet in
space-time. In the form originally advocated by Nambu [18] and Goto [19],
the action for a string is simply proportional to the area of its world
sheet. Mathematically, one formula for the area of a sheet embedded in
space-time is 
\begin{equation}
S=-T\int d\tau d\sigma \sqrt{-g}  \tag{5.1}
\end{equation}
where 
\begin{equation}
g=\det g_{ab}  \tag{5.2}
\end{equation}
\begin{equation}
g_{ab}=\delta _{\mu \nu }\partial _{a}x^{\mu }\partial _{b}x^{\nu } 
\tag{5.3}
\end{equation}
in which $x^{\mu }=x^{\mu }(\tau ,\sigma ),$ $a=0,1$ and primes will denote
derivatives with respect to $\sigma $. $T$ \ is a constant of
proportionality required to make the action dimensionless. It must have
dimension of $(length)^{-2}$ or $(mass)^{2}$ and it can be shown [17] that \ 
$T$ \ is actually the tension in the string. The string tension defines an
energy scale in string theory, and the high energy limit of the theory
corresponds to a vanishing tension value [6]. This limit is expected to
describe the string dynamics at the Planck length [6].

It is difficult to work with action (5.1) because it is highly non-linear.
An equivalent, but more convenient form of the action can be written if we
introduce a new variable $h_{ab}$, which will be a metric tensor for the
string world sheet geometry. This more convenient action is [20] 
\begin{equation}
S=-\frac{T}{2}\int d\tau d\sigma \sqrt{-h}h^{ab}\delta _{\mu \nu }\partial
_{a}x^{\mu }\partial _{b}x^{\nu }  \tag{5.4}
\end{equation}
Action (5.4) is the standard form for coupling $D$ massless scalar fields $%
x^{\mu }$ to (1+1)-dimensional gravity [21]. Since the derivatives of $%
h_{ab} $ do not appear in action (5.4) its equation of motion is a
constraint and $h_{ab}$ can be integrated out, giving back action (5.1).

Actions (5.1) and (5.4) are invariant under general transformations of the
world sheet coordinates, $\tau \rightarrow \tau ^{\prime }(\tau ,\sigma ),$ $%
\sigma \rightarrow \sigma ^{\prime }(\tau ,\sigma ).$ This reparametrization
invariance is essential for solving the classical equation of motion for $%
x^{\mu }$ that follows from action (5.4). The symmetric tensor $h_{ab}$ has
three independent components, and by a suitable choice of new parameters, $%
\tau ^{\prime }$ and $\sigma ^{\prime },$ one can gauge away two of these
components. This leaves only one independent component. However, there is
one more local two-dimensional symmetry of the string action (5.4). There is
a local Weyl scaling of the metric 
\begin{equation}
h_{ab}\rightarrow \Lambda (\tau ,\sigma )h_{ab}  \tag{5.5}
\end{equation}
which leaves the factor $\sqrt{-h}h^{ab}$ invariant. The reparametrization
invariance together with the Weyl scaling can then be used to gauge away all
the three independent components of $h_{ab}$ by imposing the conformal gauge 
$h_{ab}=\eta _{ab}$, where $\eta _{ab}$ \ is the flat two-dimensional
metric. In this gauge action (5.4) becomes 
\begin{equation}
S=-\frac{T}{2}\int d\tau d\sigma \eta ^{ab}\delta _{\mu \nu }\partial
_{a}x^{\mu }\partial _{b}x^{\nu }  \tag{5.6}
\end{equation}
The equation of motion derived from the gauge-fixed action (5.6) is the free
two-dimensional wave equation 
\begin{equation}
\frac{\partial ^{2}x^{\mu }}{\partial \tau ^{2}}-\frac{\partial ^{2}x^{\mu }%
}{\partial \sigma ^{2}}=0  \tag{5.7}
\end{equation}
There are two types of solution for this wave equation, corresponding to
open strings and closed strings. For open strings the solution must satisfy $%
\acute{x}^{\mu }=0$ at the string end points $\sigma =0$ and $\sigma =\pi $.
This solution is 
\begin{equation}
x^{\mu }(\tau ,\sigma )=x_{0}^{\mu }+p_{0}^{\mu }\tau +i\sum_{n\neq 0}\frac{1%
}{n}\alpha _{n}^{\mu }e^{-in\tau }\cos n\sigma  \tag{5.8}
\end{equation}
where $x_{0}^{\mu }$ and $p_{0}^{\mu }$ are the center of mass position and
momentum.

For closed strings the solution must satisfy the periodic boundary condition 
$x^{\mu }(\tau ,\sigma )=x^{\mu }(\tau ,\sigma +\pi )$. The solution can
then be decomposed into right-moving waves and left-moving waves, 
\begin{equation}
x^{\mu }(\tau ,\sigma )=x_{R}^{\mu }(\tau -\sigma )+x_{L}^{\mu }(\tau
+\sigma )  \tag{5.9}
\end{equation}
\begin{equation}
x_{R}^{\mu }=\frac{1}{2}x_{0}^{\mu }+\frac{1}{2}p_{0}^{\mu }(\tau -\sigma )+%
\frac{i}{2}\sum_{n\neq 0}\frac{1}{n}\alpha _{n}^{\mu }e^{-2in(\tau -\sigma )}
\tag{5.10}
\end{equation}
\begin{equation}
x_{L}^{\mu }=\frac{1}{2}x_{0}^{\mu }+\frac{1}{2}p_{0}^{\mu }(\tau +\sigma )+%
\frac{i}{2}\sum_{n\neq 0}\frac{1}{n}\tilde{\alpha}_{n}^{\mu }e^{-2in(\tau
+\sigma )}  \tag{5.11}
\end{equation}
The open string solution (5.8) played a central role in the developments of
[8] and in a wider perspective solutions (5.8-5.11) form the basis of the
most traditional quantization procedure in bosonic string theory [17]. But
there is an inconvenience in this quantization procedure: the Weyl
invariance (5.5) is not preserved in the quantized theory. Only in (25+1)
space-time dimensions can we construct a consistent quantized bosonic string
theory based on the conformal wave equation (5.7) because only in this
space-time dimension is the Weyl invariance (5.5) restored in the quantum
theory. This situation could in principle be avoided if we could find a way
to the wave equation (5.7) that does not require the Weyl invariance of the
theory. We now show that we can find this way by treating the string as a
constrained system, in much the same way as we did for the relativistic
particle. Following this way, we will also be able to show that the basic
commutator structure we found for the massless particle can also be
encountered in the high energy limit of string theory.

In the transition to the Hamiltonian formalism the Nambu-Goto action (5.1)
gives the canonical momentum 
\begin{equation}
p_{\mu }=-T\sqrt{-g}g^{0a}\partial _{a}x_{\mu }  \tag{5.12}
\end{equation}
and this momentum gives rise to the primary constraints 
\begin{equation}
\Phi _{0}=\frac{1}{2}(p^{2}+T^{2}\acute{x}^{2})=0  \tag{5.13}
\end{equation}
\begin{equation}
\Phi _{1}=p.\acute{x}=0  \tag{5.14}
\end{equation}
The canonical Hamiltonian corresponding to action (5.1) identically vanishes
and so we can construct the first-order Lagrangian density 
\begin{equation}
L=p.\dot{x}-\frac{\lambda _{0}}{2}(p^{2}+T^{2}\acute{x}^{2})-\lambda _{1}p.%
\acute{x}  \tag{5.15}
\end{equation}
where the two-dimensional fields $\lambda _{0}$ and $\lambda _{1}$ play the
role of Lagrange multipliers. Solving the equation of motion for $p_{\mu }$
that follows from (5.15) and inserting the result back in it, we obtain the
string action 
\begin{equation}
S=\int d\tau d\sigma \lbrack \frac{1}{2}\lambda _{0}^{-1}(\dot{x}-\lambda
_{1}\acute{x})^{2}-\frac{1}{2}\lambda _{0}T^{2}\acute{x}^{2}]  \tag{5.16}
\end{equation}
Action (5.16) is the higher-dimensional extension of the particle action
(3.6). Because the constraints $\Phi _{0}$ and $\Phi _{1}$ are first-class,
now we have two arbitrary functions, $\lambda _{0}$ and $\lambda _{1}$, at
our disposal. Under infinitesimal reparametrizations we have 
\begin{equation}
\delta (\partial _{a}x^{\mu })=-\partial _{a}\epsilon ^{b}\partial
_{b}x^{\mu }  \tag{5.17}
\end{equation}
and action (5.16) will be reparametrization-invariant if 
\begin{equation}
\delta \lambda _{0}=(\acute{\epsilon}_{1}-\dot{\epsilon}_{0})\lambda
_{0}-\lambda _{1}\lambda _{0}\acute{\epsilon}_{0}  \tag{5.18a}
\end{equation}
\begin{equation}
\delta \lambda _{1}=(\acute{\epsilon}_{1}-\dot{\epsilon}_{0})\lambda _{1}+%
\dot{\epsilon}_{1}-\acute{\epsilon}_{0}\lambda _{1}^{2}-T^{2}\lambda _{0}^{2}%
\acute{\epsilon}_{0}  \tag{5.18b}
\end{equation}
The string action (5.16) then has the usual reparametrization invariance of
the Nambu-Goto action. The classical equation of motion for $x^{\mu }$ that
follows from action (5.16) is 
\begin{equation}
\frac{\partial }{\partial \tau }(\frac{1}{\lambda _{0}}\dot{x}^{\mu }-\frac{%
\lambda _{1}}{\lambda _{0}}\acute{x}^{\mu })+\frac{\partial }{\partial
\sigma }[-\frac{\lambda _{1}}{\lambda _{0}}\dot{x}^{\mu }+(\frac{\lambda
_{1}^{2}}{\lambda _{0}}-\lambda _{0}T^{2})\acute{x}^{\mu }]=0  \tag{5.19}
\end{equation}
Choosing first $\lambda _{0}=1$, if we want to reach the conformal wave
equation, we must satisfy the condition $\lambda _{1}^{2}-T^{2}=-1$. This
condition means that $\lambda _{1}=\pm \sqrt{T^{2}-1}$. Using $\lambda
_{0}=1 $, the positive value of $\lambda _{1}$ in the first term of (5.19),
and the negative value of $\lambda _{1}$ in the second term, equation (5.19)
becomes 
\begin{equation*}
\frac{\partial ^{2}x^{\mu }}{\partial \tau ^{2}}-\frac{\partial ^{2}x^{\mu }%
}{\partial \sigma ^{2}}=0
\end{equation*}
which is identical to the conformal wave equation (5.7), obtained from
action (5.4) after imposing the Weyl invariance of the theory.

Returning to the string action (5.16), we now use the reparametrization
invariance to impose the gauge $\lambda _{1}=0$. It can be verified that
constraints (5.13) and (5.14) are preserved in this gauge. Choosing again $%
\lambda _{0}=1$, we find that in the energy scale where the string tension
has an unitary value, the resulting gauge-fixed action can be written as 
\begin{equation}
S=\int d\tau d\sigma (\frac{1}{2}\sqrt{-\det \mathbf{h}}\mathbf{h}^{-1}%
\mathbf{Trg)}  \tag{5.20}
\end{equation}
where the matrix fields 
\begin{equation}
\mathbf{h}=\left[ 
\begin{array}{cc}
1 & 0 \\ 
0 & -1
\end{array}
\right]  \tag{5.21}
\end{equation}
\begin{equation}
\mathbf{g=}\left[ 
\begin{array}{cc}
\partial _{0}x.\partial _{0}x & \partial _{0}x.\partial _{1}x \\ 
\partial _{1}x.\partial _{0}x & \partial _{1}x.\partial _{1}x
\end{array}
\right]  \tag{5.22}
\end{equation}
were introduced. Action (5.20) is in the usual form of a non-commutative
field theory action [10]. It can be checked that the conformal wave equation
(5.7) directly follows from an action principle based in (5.20).

Returning once more to our starting string action (5.16), we choose again $%
\lambda _{1}=0$ but now select the energy scale where the string tension
vanishes. The string action in this limit becomes 
\begin{equation}
S=\frac{1}{2}\int d\tau d\sigma \lambda _{0}^{-1}\dot{x}^{2}  \tag{5.23}
\end{equation}
and to be fully consistent, must be complemented with the constraints 
\begin{equation}
\Phi _{0}=\frac{1}{2}p^{2}=0  \tag{5.24}
\end{equation}
\begin{equation}
\Phi _{1}=p.\acute{x}=0  \tag{5.25}
\end{equation}
which are the $T=0$ limit of constraints (5.13-5.14). The equation of motion
for $x^{\mu }$ that follows from the tensionless string action (5.23) is
identical in form to the equation of motion for $x^{\mu }$ that follows from
the massless particle action (3.8) (with the total $\tau $-derivative
replaced by a partial $\tau $-derivative). We see that in the high energy
limit there is a string motion in which each point of the string moves as a
massless particle. As a consequence, the string action (5.23) is invariant
under the scale transformation (3.11) and under the conformal transformation
(3.12). Relaxing again the mass-shell condition (5.24) we find that the
tensionless action (5.23) is also invariant under the transformation (3.13),
the only difference being that now $x^{\mu }=x^{\mu }(\tau ,\sigma )$. The
commutator structure (3.20-3.22) has then a natural extension in the high
energy limit of bosonic string theory. This extension is simply given by 
\begin{equation}
\lbrack \tilde{p}_{\mu }(\sigma ),\tilde{p}_{\nu }(\sigma ^{\prime })]=0 
\tag{5.26}
\end{equation}
\begin{equation}
\lbrack \tilde{x}_{\mu }(\sigma ),\tilde{p}_{\nu }(\sigma ^{\prime
})]=\{i\delta _{\mu \nu }(1+\beta )^{2}+(1+\beta )[x_{\mu },\beta ]p_{\nu
}\}\delta (\sigma -\sigma ^{\prime })  \tag{5.27}
\end{equation}
\begin{equation}
\lbrack \tilde{x}_{\mu }(\sigma ),\tilde{x}_{\nu }(\sigma ^{\prime
})]=(1+\beta )\{x_{\mu }[\beta ,x_{\nu }]-x_{\nu }[\beta ,x_{\mu }]\}\delta
(\sigma -\sigma ^{\prime })  \tag{5.28}
\end{equation}
\bigskip

\section{Superstrings}

We now explicitly check that the bosonic commutators (5.26-5.28) remain
invariant in the high energy limit of superstring theory. The Green-Schwarz
space-time supersymmetric version of the Nambu-Goto string can be written as 
\begin{equation}
S=-T\int d\tau d\sigma (\sqrt{-g}+L_{WZ})  \tag{6.1}
\end{equation}
where $g$ is the determinant of the induced two-dimensional metric $%
g_{ab}=\delta _{\mu \nu }Z_{a}^{\mu }Z_{b}^{\nu }$ with $Z_{a}^{\mu
}=\partial _{a}X^{\mu }-i\bar{\theta}\Gamma ^{\mu }\partial _{a}\theta $. $%
L_{WZ}$ is the Wess-Zumino term 
\begin{equation}
L_{WZ}=-i\epsilon ^{ab}Z_{a}.\bar{\theta}\Gamma \partial _{b}\theta 
\tag{6.2}
\end{equation}
which is necessary for the presence of the local fermionic symmetry [17].
This symmetry is important at the classical level, but we will present
evidence that it may be discarded in the quantized theory. Switching off the
fermions in action (6.1) we recover the original Nambu-Goto action (5.1) for
the bosonic string. At this point it should be clear to the reader that we
can immediately construct the supersymmetric extension of the bosonic string
action (5.16) by simply making the substitution $\partial _{a}x^{\mu
}\rightarrow Z_{a}^{\mu }$ and adding the Wess-Zumino term. But let us check
this explicitly.

The canonical momenta that follow from action (6.1) are 
\begin{equation}
P_{\mu }=-T(\sqrt{-g}g^{0a}Z_{a\mu }+i\bar{\theta}\Gamma _{\mu }\dot{\theta})
\tag{6.3}
\end{equation}
\begin{equation}
\Pi _{\alpha }=i(P-TZ_{1}).(\Gamma \theta )_{\alpha }  \tag{6.4}
\end{equation}
and these momenta give rise to the primary constraints 
\begin{equation}
\Phi _{0}=\frac{1}{2}(Q^{2}+T^{2}Z_{1}^{2})=0  \tag{6.5}
\end{equation}
\begin{equation}
\Phi _{1}=Q.Z_{1}=0  \tag{6.6}
\end{equation}
\begin{equation}
\Psi _{\alpha }=\Pi _{\alpha }-i(P-TZ_{1}).(\Gamma \theta )_{\alpha }=0 
\tag{6.7}
\end{equation}
where we introduced the mechanical momentum 
\begin{equation}
Q_{\mu }=P_{\mu }+iT\bar{\theta}\Gamma _{\mu }\acute{\theta}  \tag{6.8}
\end{equation}

The canonical Hamiltonian corresponding to action (6.1) identically vanishes
and so Dirac's Hamiltonian density for the Nambu-Goto superstring is 
\begin{equation}
H_{D}=\lambda ^{a}\Phi _{a}+\bar{\varsigma}\Psi  \tag{6.9}
\end{equation}
where $\lambda ^{a}$ are two bosonic Lagrange multipliers and $\bar{\varsigma%
}$ is a fermionic one. The Lagrangian density corresponding to (6.9) is 
\begin{equation}
L=P.\dot{X}+\bar{\Pi}\dot{\theta}-\lambda ^{a}\Phi _{a}-\bar{\varsigma}\Psi 
\tag{6.10}
\end{equation}
If we solve the classical equations of motion for $P_{\mu }$ and $\bar{\Pi}%
_{\alpha }$ that follow from an action principle based in (6.10), and insert
the results back in it, we obtain the superstring action 
\begin{equation}
S=\int d\tau d\sigma \lbrack \frac{1}{2}\lambda _{0}^{-1}(Z_{0}-\lambda
^{1}Z_{1})^{2}-\frac{1}{2}\lambda ^{0}T^{2}Z_{1}^{2}+TL_{WZ}]  \tag{6.11}
\end{equation}
Action (6.11) is the supersymmetric extension of the bosonic string action
(5.16). In fact, under the infinitesimal reparametrizations 
\begin{equation}
\delta (\partial _{a}X^{\mu })=-\partial _{a}\epsilon ^{b}\partial
_{b}X^{\mu }  \tag{6.12a}
\end{equation}
\begin{equation}
\delta (\partial _{a}\theta _{\alpha })=-\partial _{a}\epsilon ^{b}\partial
_{b}\theta _{\alpha }  \tag{6.12b}
\end{equation}
the Wess-Zumino term is well-known to be invariant [17], and action (6.11)
becomes reparametrization-invariant if $\lambda _{0}$ varies as in equation
(5.18a) and $\lambda _{1}$ as in equation (5.18b). We can again impose the
gauge $\lambda _{0}=1,$ \ $\lambda _{1}=0$ and in the energy scale where the
string tension has an unitary value we may write the non-commutative
supersymmetric action 
\begin{equation}
S=\int d\tau d\sigma (\frac{1}{2}\sqrt{-\det \mathbf{h}}\mathbf{h}^{-1}%
\mathbf{Trg+}L_{WZ})  \tag{6.13}
\end{equation}
where now 
\begin{equation}
\mathbf{g=}\left[ 
\begin{array}{cc}
Z_{0}.Z_{0} & Z_{0}.Z_{1} \\ 
Z_{1}.Z_{0} & Z_{1.}Z_{1}
\end{array}
\right]  \tag{6.14}
\end{equation}
It is interesting to discard the Wess-Zumino term in action (6.13). This is
equivalent to relaxing the fermionic constraints (6.7) and being off shell
in the fermionic variables. Dropping $L_{WZ}$ and varying $X_{\mu }$ in
action (6.13) we arrive at the equations 
\begin{equation}
\frac{\partial ^{2}X^{\mu }}{\partial \tau ^{2}}-\frac{\partial ^{2}X^{\mu }%
}{\partial \sigma ^{2}}=0  \tag{6.15a}
\end{equation}
\begin{equation}
\frac{\partial ^{2}\theta _{\alpha }}{\partial \tau ^{2}}-\frac{\partial
^{2}\theta _{\alpha }}{\partial \sigma ^{2}}=0  \tag{6.15b}
\end{equation}
The same set of equations is arrived at by varying $\theta _{\alpha }$ in
action (6.13) without the $L_{WZ}$ term. The set (6.15) is the
supersymmetric extension of the conformal equation (5.7). \textsl{\ }

Obviously we may again impose the necessary conditions to arrive at the
supersymmetric extension of the tensionless bosonic string action (5.23).
This extension is given by the action 
\begin{equation}
S=\frac{1}{2}\int d\tau d\sigma \lambda ^{-1}Z_{0}^{2}  \tag{6.16}
\end{equation}
complemented with the constraints 
\begin{equation}
\Phi _{0}=\frac{1}{2}P^{2}=0  \tag{6.17}
\end{equation}
\begin{equation}
\Phi _{1}=P.Z_{1}=0  \tag{6.18}
\end{equation}
\begin{equation}
\Psi _{\alpha }=\Pi _{\alpha }-iP.\Gamma \theta _{\alpha }=0  \tag{6.19}
\end{equation}
From our\ previous experience with the massless superparticle we note that
the bosonic commutators (5.26-5.28) must be preserved in the superstring
theory defined by action (6.16).

\section{Relativistic membranes}

As a final task, we now briefly show that the basic bosonic commutator
structure we found for the massless particle and for the tensionless string
also exists in the tensionless limit of relativistic membrane theory. For a
review of membrane theory see [22].

A relativistic membrane propagating in space-time may be described by a
Nambu-Goto-Dirac-type action given by 
\begin{equation}
S=-T\int d\tau d^{2}\sigma \sqrt{-g}  \tag{7.1}
\end{equation}
with 
\begin{equation}
g=\det g_{AB}  \tag{7.2}
\end{equation}
\begin{equation}
g_{AB}=\delta _{\mu \nu }\partial _{A}X^{\mu }\partial _{B}X^{\nu } 
\tag{7.3}
\end{equation}
where now $X^{\mu }=X^{\mu }(\tau ,\sigma _{1},\sigma _{2})$ and $A=0,1,2$.
Action (7.1) gives the canonical momentum 
\begin{equation}
P_{\mu }=-T\sqrt{-g}g^{0A}\partial _{A}X_{\mu }  \tag{7.4}
\end{equation}
and this momentum gives rise to the primary constraints 
\begin{equation}
\Phi _{0}=\frac{1}{2}(P^{2}+T^{2}\tilde{g})=0  \tag{7.5}
\end{equation}
\begin{equation}
\Phi _{a}=P.\partial _{a}X=0  \tag{7.6}
\end{equation}
where $\tilde{g}=\det (\partial _{a}X.\partial _{b}X)$ and $a=1,2$.
Performing the same manipulations as in the bosonic string, we arrive at the
membrane action 
\begin{equation}
S=\int d\tau d^{2}\sigma \lbrack \frac{1}{2}\lambda _{0}^{-1}(\dot{X}%
-\lambda ^{a}\partial _{a}X)^{2}-\frac{1}{2}\lambda _{0}T^{2}\tilde{g}] 
\tag{7.7}
\end{equation}
Action (7.7) is the higher-dimensional extension of the string action (5.16)
It is reparametrization invariant for 
\begin{equation}
\delta \lambda ^{0}=(\partial _{a}\epsilon ^{a}-\partial _{0}\epsilon
^{0})\lambda ^{0}-2\lambda ^{0}\lambda ^{a}\partial _{a}\epsilon ^{0} 
\tag{7.8a}
\end{equation}
\begin{equation}
\delta \lambda ^{a}=(\partial _{0}+\lambda ^{b}\partial _{b})\epsilon
^{a}-\lambda ^{a}(\partial _{0}+\lambda ^{b}\partial _{b})\epsilon
^{0}-(\lambda ^{0})^{2}T^{2}\tilde{g}\tilde{g}^{ab}\partial _{b}\epsilon ^{0}
\tag{7.8b}
\end{equation}
Equations (7.8b) show that there is enough reparametrization freedom to
impose the gauge $\lambda _{1}=\lambda _{2}=0.$ In this gauge we get the
membrane action 
\begin{equation}
S=\int d\tau d^{2}\sigma (\frac{1}{2}\lambda _{0}^{-1}\dot{X}^{2}-\frac{1}{2}%
\lambda _{0}T^{2}\tilde{g})  \tag{7.9}
\end{equation}
It is easy to check that action (7.9) preserves the constraints (7.5) and
(7.6). Going to the energy region where the membrane tension vanishes, the
partially gauge-fixed action (7.9) becomes 
\begin{equation}
S=\frac{1}{2}\int d\tau d^{2}\sigma \lambda _{0}^{-1}\dot{X}^{2}  \tag{7.10}
\end{equation}
complemented with the constraints 
\begin{equation}
\Phi _{0}=\frac{1}{2}P^{2}=0  \tag{7.11}
\end{equation}
\begin{equation}
\Phi _{a}=P.\partial _{a}X=0  \tag{7.12}
\end{equation}
We see that each point of this partially gauge-fixed tensionless membrane
moves as a massless particle, and each line moves as a tensionless string.
As a consequence of this, action (7.10) is invariant under the scale
transformation (3.11), under the conformal transformation (3.12) and also
under the transformation (3.13), now with $X^{\mu }=X^{\mu }(\tau ,\sigma
_{1},\sigma _{2})$. The commutator structure (5.26-5.28) may then be further
extended to 
\begin{equation}
\lbrack \tilde{P}_{\mu }(\sigma _{1,}\sigma _{2}),\tilde{P}_{\nu }(\sigma
_{1}^{\prime },\sigma _{2}^{\prime })]=0  \tag{7.13}
\end{equation}
\begin{equation*}
\lbrack \tilde{X}_{\mu }(\sigma _{1},\sigma _{2}),\tilde{P}_{\nu }(\sigma
_{1}^{\prime },\sigma _{2}^{\prime })]=\{i\delta _{\mu \nu }(1+\beta )^{2}.
\end{equation*}
\begin{equation}
+(1+\beta )[X_{\mu },\beta ]P_{\nu }\}\delta (\sigma _{1}-\sigma
_{1}^{\prime })\delta (\sigma _{2}-\sigma _{2}^{\prime })  \tag{7.14}
\end{equation}
\begin{equation*}
\lbrack \tilde{X}_{\mu }(\sigma _{1},\sigma _{2}),\tilde{X}_{\nu }(\sigma
_{1}^{\prime },\sigma _{2}^{\prime })]=(1+\beta )\{X_{\mu }[\beta ,X_{\nu }]
\end{equation*}
\begin{equation}
-X_{\nu }[\beta ,X_{\mu }]\}\delta (\sigma _{1}-\sigma _{1}^{\prime })\delta
(\sigma _{2}-\sigma _{2}^{\prime })  \tag{7.15}
\end{equation}
The space-time supersymmetric extension of action (7.7) may now be
immediately written, 
\begin{equation}
S=\int d\tau d^{2}\sigma \lbrack \frac{1}{2}\lambda _{0}^{-1}(Z_{0}-\lambda
^{a}Z_{a})^{2}-\frac{1}{2}\lambda _{0}T^{2}\tilde{G}+TL_{WZ}]  \tag{7.16}
\end{equation}
where now $\tilde{G}=\det Z_{a}.Z_{b},$ and the Wess-Zumino term is 
\begin{equation}
L_{WZ}=-i\epsilon ^{ABC}\bar{\theta}\Gamma _{\mu \nu }\partial _{A}\theta
(Z_{B}^{\mu }\partial _{C}X^{\nu }-\frac{1}{3}\bar{\theta}\Gamma ^{\mu
}\partial _{B}\theta \bar{\theta}\Gamma ^{\nu }\partial _{C}\theta ) 
\tag{7.17}
\end{equation}
with $\Gamma _{\mu \nu }=\Gamma _{\lbrack \mu }\Gamma _{\nu ]}$ . It is
clear that the supersymmetric extension of the tensionless membrane action
(7.10) can be immediately constructed, and its invariance under the membrane
extension of the massless superparticle transformation (4.8) will lead to
bosonic membrane canonical variables satisfying commutators (7.13-7.15).

\section{Concluding remarks}

In this work we studied relativistic particle, string and membrane theories
as defining field theories containing gravity in (0+1), (1+1) and (2+1)
space-time dimensions, respectively. \ The massless limit of particle
theory, and the tensionless limit of string and membrane theories were
investigated here using alternative Lagrangian formulations obtained by
incorporating the Hamiltonian constraints into the formalism. These limits
were interpreted as describing the high energy limits of the corresponding
theories. We showed how we may use the ortogonality condition between the
relativistic velocity and the relativistic acceleration, a condition which
in a certain sense defines the region of applicability of Special
Relativity, to induce the appearance of a new invariance of the massless
particle action. It was then described how this invariance allows the
construction of an off shell extension of the conformal algebra and induces
a transition to new space-time coordinates which obey non-vanishing
commutation relations. These commutation relations are different from the
ones usually encountered in the modern formulations of non-commutative field
theory and suggest a dynamical, rather than static, space-time geometry.
These bosonic commutation relations remain unaltered for the massless
supersymmetric particle action. It was then shown how extensions of these
non-vanishing bosonic commutation relations can also be encountered in the
tensionless limit of string and membrane theory, both in the bosonic and
supersymmetric sectors.

Using special-relativistic concepts only, we showed how we may construct a
simple model of a relativistic particle with an internal structure, an idea
originally developed by Dirac [23] in 1971. We then proposed that the
non-locality introduced by the commutators we found for the massless
particle should be interpreted as an indication that at very high energies
the massless particle could behave as a gravitational dipole. The presence
of the same basic commutator structure in the tensionless limit of string
and membrane theories should then be interpreted as evidences that there are
possible motions in which these relativistic objects behave as linear and
surface distributions of gravitational dipoles, respectively. What role
would the concept of a gravitational dipole play in Physics?

A gravitational dipole has a dynamical behavior much different from that of
an electric dipole. While opposite electric charges attract themselves, it
is seen from Newton's equation for the gravitational force 
\begin{equation}
\vec{F}=-\frac{\gamma m_{1}m_{2}}{r^{2}}\vec{u}_{r}  \tag{8.1}
\end{equation}
that opposite masses should repel each other. Then, contrary to electric
dipoles which require an external electric field to prevent them from
collapsing, gravitational dipoles may exist indefinitely as consequences of
their own internal repulsive gravitational forces.

Newton's theory for the gravitational field $\vec{G}$ is contained in the
equations 
\begin{equation}
\vec{\nabla}.\vec{G}=4\pi \gamma \rho  \tag{8.2a}
\end{equation}
\begin{equation}
\vec{\nabla}\times \vec{G}=0  \tag{8.2b}
\end{equation}
Equations (8.2) are identical in form to Maxwell's equations in the absence
of electric currents and magnetic fields 
\begin{equation}
\vec{\nabla}.\vec{E}=\frac{\rho }{\epsilon _{0}}  \tag{8.3a}
\end{equation}
\begin{equation}
\vec{\nabla}\times \vec{E}=0  \tag{8.3b}
\end{equation}
It is well-known that equations (8.3) are just a part of a complete set of
equations, the Maxwell equations 
\begin{equation}
\vec{\nabla}.\vec{E}=\frac{\rho }{\epsilon _{0}}  \tag{8.4a}
\end{equation}
\begin{equation}
\vec{\nabla}\times \vec{E}=-\frac{\partial \vec{B}}{\partial t}  \tag{8.4b}
\end{equation}
\begin{equation}
\vec{\nabla}.\vec{B}=0  \tag{8.4c}
\end{equation}
\begin{equation}
\vec{\nabla}\times \vec{B}=\mu _{0}\vec{J}+\epsilon _{0}\mu _{0}\frac{%
\partial \vec{E}}{\partial t}  \tag{8.4d}
\end{equation}
and that this set of four equations can be naturally written in covariant
form with the introduction of a four potential $A^{\mu }=(\phi ,\vec{A})$.

Now, a stationary point mass constitutes a gravitational charge only for
observers at rest relative to it or, at most, for observers moving with very
small velocities. For fast-moving observers a gravitational charge looks
like a gravitational current. Gravitational currents are then necessary
ingredients to achieve covariance. It now seems that, to get a covariant
Newtonian gravitation, it is only necessary to have a gravitational analogue
of the magnetic field $\vec{B}$ , a field, say $\vec{M}$ , satisfying 
\begin{equation}
\vec{\nabla}.\vec{M}=0  \tag{8.5a}
\end{equation}
\begin{equation}
\vec{M}=\vec{\nabla}\times \vec{W}  \tag{8.5b}
\end{equation}
for some vector potential $\vec{W}$ . Perhaps the gravitational dipoles
could be used as sources for the field $\vec{M}$ , as magnetic dipoles are
sources for the field $\vec{B}$ . As the magnetic field $\vec{B}$ is a
manifestation of electric charges in motion, the field $\vec{M}$ would be a
manifestation of gravitational charges in motion, and this is reflected in
the fact that the gravitational dipole picture can only be constructed in
the relativistic theory. It may be interesting to investigate the internal
consistency of such a complemented Newtonian gravitational field theory. It
would describe attractive and repulsive gravitational interactions,
intermediated by spin-one gauge particles, in a flat universe which would be
naturally expanding due to the presence of negative matter.

This work is dedicated to the $100^{th}$ anniversary of Special Relativity
(1905-2005).

\end{document}